\documentstyle[12pt]{article}
\input psfig    

\renewcommand{\thefootnote}{\fnsymbol{footnote}}	

\newcommand{\lesssim}{\mathop{}_{\textstyle \sim}^{\textstyle <}}
\newcommand{\gtrsim}{\mathop{}_{\textstyle \sim}^{\textstyle >}}
\newcommand{\text}{\mbox}
\newcommand{\ser}{\tilde{e}_R}
\newcommand{\sel}{\tilde{e}_L}
\newcommand{\epem}{e^+e^-}
\newcommand{\emem}{e^-e^-}
\newcommand{\neutralino}{\tilde{\chi}^0}
\newcommand{\LSP}{\tilde{\chi}^0_1}
\newcommand{\mser}{m_{\tilde{e}_R}}
\newcommand{\mLSP}{m_{\tilde{\chi}^0_1}}

\begin{document}







\begin{titlepage}
\begin{center}

\hfill    FERMILAB-CONF-98/005-T\\
\hfill    hep-ph/9801234 \\
\hfill    January 1998

\vskip .5in

{ \Large \bf Precision Supersymmetry Measurements \\
at the $e^-e^-$ Collider } 

\vskip .3in

{\large Hsin-Chia Cheng}

\vskip .2in

{\it Fermi National Accelerator Laboratory\\
     P.O. Box 500\\
     Batavia, IL 60510}

\end{center}

\vskip .2in

\begin{abstract}
Measurements of supersymmetric particle couplings provide important
verification of supersymmetry. If some of the superpartners are
at the multi-TeV scale, they will escape direct detection at planned 
future colliders. However, such particles induce nondecoupling 
corrections in processes involving the accessible superparticles 
through violations of the supersymmetric equivalence between 
gauge boson and gaugino couplings. These violations are analogous 
to the oblique corrections in the electroweak sector of the standard 
model, and can be parametrized in terms of super-oblique parameters. 
The $e^- e^-$ collision mode of a future linear collider is shown
to be an excellent environment for such high precision measurements
of these SUSY parameters, which will provide an important probe
of superparticles beyond reachable energies.
\end{abstract}
\vskip .3in

\begin{center}
Talk presented at the 2nd International Workshop on 
Electron-Electron Interactions at TeV Energies, 
September 22-24, 1997, University of California, Santa Cruz

\end{center}
\end{titlepage}

\renewcommand{\thepage}{\arabic{page}}
\setcounter{page}{1}
\renewcommand{\thefootnote}{\arabic{footnote}}
\setcounter{footnote}{0}


\section{Introduction}

If supersymmetry (SUSY) is relevant for the hierarchy problem,
the supersymmetric partners of ordinary particles should have
masses on the order of TeV scale. The discovery of supersymmetric
particles at the present and future colliders is therefore
promising. After the discovery, measurements of the superparticle
properties such as their masses and couplings, will be the focus
of studies. In particular, we need to check whether these new
particles are indeed the superpartners of the Standard Model (SM)
particles. This can be done by measuring the couplings of the 
superparticles, which is related to the couplings of the SM particles
by supersymmetry. For example, SUSY implies the relations
\begin{equation}
\label{susyrel}
g_i=h_i ,
\end{equation}
where $g_i$ are the SM gauge couplings, $h_i$ are their SUSY
analogues, the gaugino-fermion-sfermion couplings, and the 
subscript $i=1, \, 2, \, 3$ refers to the U(1), SU(2), and 
SU(3) gauge groups, respectively. Unlike other relations,
such as the unification of gaugino masses, these relations
hold in all SUSY models and are true to all orders in the limit
of unbroken SUSY. Therefore, they can serve as robust tests
of SUSY. Such tests of SUSY were first considered by Feng et al
in Ref.~1, in which the chargino production at 
a future $e^+ e^-$ linear collider is used. 
Tests with slepton production
at an $e^+ e^-$ linear collider is then considered by 
Nojiri et al~\cite{Nojiri}.

Because SUSY is broken, the relation (\ref{susyrel}) will
receive radiative corrections due to SUSY breaking. Especially
when some of the superpartners have large SUSY breaking masses,
the deviations from Eq.~(\ref{susyrel}) can be significant and
may be used to probe SUSY breaking mass splittings. In fact,
there are many models in which some number of the superpartners
of ordinary matter and gaugino fields are very heavy and may be
beyond the discovery range of the planned future colliders.
These heavy particles decouple from most low energy processes.
However, because of their large SUSY breaking masses, at the
lower energy scale of the light superpartners, they induce
deviations from the SUSY relations Eq.~(\ref{susyrel}) through
radiative corrections. The deviations grow logarithmically
as the heavy superpartner masses increase. Therefore, if the
gaugino couplings can be measured to the precision sensitive
to the typical deviation of (\ref{susyrel}) from such a heavy
superpartner sector, not only can we test SUSY, but also probe
the scale of the heavy superpartners. This can help to set the
target energies of future colliders in searching for these particles.
As we will see, the $e^- e^-$ collider is an excellent tool
for the precision measurements of some of the superparticle couplings.

\section{Super-oblique corrections}

Before discussing the measurements at the $e^- e^-$ collider, we
first discuss what kind of precision we would like to achieve,
in order to be sensitive to the heavy superpartner scale.
The models with some heavy superpartners can be roughly
divided into two categories. In the first class of models, which we
will refer to as ``heavy QCD models,'' the gluino and all the squarks
are heavy.  Examples of such models include the no-scale 
supergravity~\cite{noscale}, models with gauge-mediated SUSY
breaking~\cite{gm}, where strongly-interacting superparticles get large
contributions to their masses,  and
models with a heavy gluino, which gives large contributions to the squark
masses through renormalization group evolution.  In the second class
of models, ``2--1 models,'' the first two generation sfermions
masses are heavy (${\cal O}(10 \mbox{ TeV})$), while the third generation
sfermions are at the weak scale~\cite{2--1}.  Such models are
motivated by the attempts to solve the SUSY flavor problem without the 
need for universality of sfermion masses or alignment, while avoiding the
extreme fine-tuning problem by keeping the third generation sfermions
which couple strongly to the Higgs sector at the weak scale.

The corrections to Eq.~(\ref{susyrel}) are very similar to the
oblique corrections of the standard model~\cite{CFP1,RKS}.  In the
standard model, nondegenerate SU(2) multiplets lead to different
renormalizations of the propagators of the $W$ and $Z$ gauge bosons,
inducing nondecoupling effects which grow with the mass splitting.
Similarly, in SUSY theories, nondegenerate supermultiplets
lead to inequivalent renormalizations of the propagators of
gauge bosons and gauginos, inducing nondecoupling
effects that grow with the mass splitting.  We will therefore refer to
the latter effects as ``super-oblique corrections'' and parametrize
them by ``super-oblique parameters''~\cite{CFP1}.  These corrections
are particularly important, because they are universal in processes
involving gauginos and enhanced by a number of factors.

The differences between the gauge couplings $g_i$ and the gaugino
couplings $h_i$ come from differences in wavefunction
renormalizations, and hence are most analogous to the oblique
parameter $U$. We therefore define
\begin{equation}
\widetilde{U}_i \equiv h_i / g_i - 1 \ ,
\end{equation}
where the subscript $i$ denotes the gauge group.
For the two categories of models discussed above, we find\cite{CFP1}
\begin{eqnarray}
\label{u1}
\widetilde{U}_1 &\approx& 0.35\% \ (0.29\%) \times \ln R \\
\label{u2}
\widetilde{U}_2 &\approx& 0.71\% \ (0.80\%) \times \ln R \\
\label{u3}
\widetilde{U}_3 &\approx& 2.5\%  \times \ln R 
\end{eqnarray}
for 2--1 models (heavy QCD models), where $R = M/m$ is the ratio of
heavy and light superpartner mass scales, and can be ${\cal O}(10)$
or even ${\cal O}(100)$ (for 2--1 models). These parameters can
also receive contributions from possible exotic supermultiplets.
For example, contributions
from vector-like (messenger) sectors have also been
calculated~\cite{CFP1}, and were found that they can be significant 
only for highly split supermultiplets with masses $\lesssim {\cal O}
(100\mbox{TeV})$.

From Eqs.~(\ref{u1})--(\ref{u3}) we see that the corrections due to
heavy superpartners can be a few percent, and are larger for
stronger couplings. In Ref.~1, from chargino production at a
linear $e^+ e^-$ collider, the SU(2) gaugino coupling $h_2$
is found to be able to be measured to ${}^{+30}_{-15}\%$ for a point
in the gaugino region. While it can serve as a test of SUSY,
it is not accurate enough to probe the super-oblique corrections.
However, if we can also measure the sneutrino masss to 1\%,
the uncertainties can be reduced to 2--3\%, starting to be 
sensitive to the heavy sector~\cite{CFP2}. In Ref.~2, the
slepton production at the linear $e^+ e^-$ collider was studied,
and they found that $h_1$ can be measured to $\sim 1\%$, which
is also about the size of the correction from a heavy sector.
However, if we really want to constrain the heavy superpartner
scale, higher precision is required. As we will see in the following,
the $e^- e^-$ option of a linear collider may give the most precise
determination of the super-oblique corrections.

\section{Measurements at the $e^-e^-$ collider}

\begin{figure}
\centerline{\psfig{file=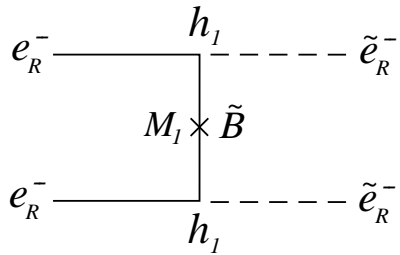,width=0.45\textwidth}}
\caption{Selectron pair production at an $e^-e^-$ collider.
\label{diagram}}
\end{figure}
The selectrons can be pair produced at an $e^- e^-$ collider
through the $t$-channel neutralino exchange (Fig.~\ref{diagram}).
Here we will focus on the U(1) couplings and consider the
$\ser$ pair production. There are several advantages with
selectron production at an $\emem$ collider.
At an $\emem$ collider, selectrons are produced only through 
$t$-channel neutralino exchange. The cross section for $\ser$
production is thus directly proportional to $h_1^4$. The coupling
$h_1$ can be extracted directly from the total cross section,
which is usually larger than at the $\epem$ collider if the Bino
mass $M_1$ is not too small. 
The backgrounds to selectron pair production at $\emem$
colliders are very small.  Most of the major backgrounds present in
the $\epem$ mode are absent; {\it e.g.}, $W$ pair and chargino pair
production are forbidden by total lepton number conservation. 
The remaining background $e^- \nu W^-$ can be suppressed by
polarizing both $e^-$ beams right-handed. In addition, because
a majorana mass insertion in the neutralino propagator is needed
to flip the chirality, the dependence of the total cross
section on $\mLSP$ is different at $\emem$ colliders from at
$\epem$ colliders. This can be exploited to reduce theoretical
systematic errors arising from uncertainties in the $\ser$ and
$\LSP$ masses. We will come to this point in more detail later.

Now let us discuss what kind of precision can be achieved
at the $\emem$ collider. As we will see, the error in the
total cross section could be well below 1\%, which means
the uncertainty in $h_1$ below 0.25\%. Such a measurement
could constrain the heavy superpartner scale to within a factor
of 3 or even better. 

\begin{figure}
\centerline{\psfig{file=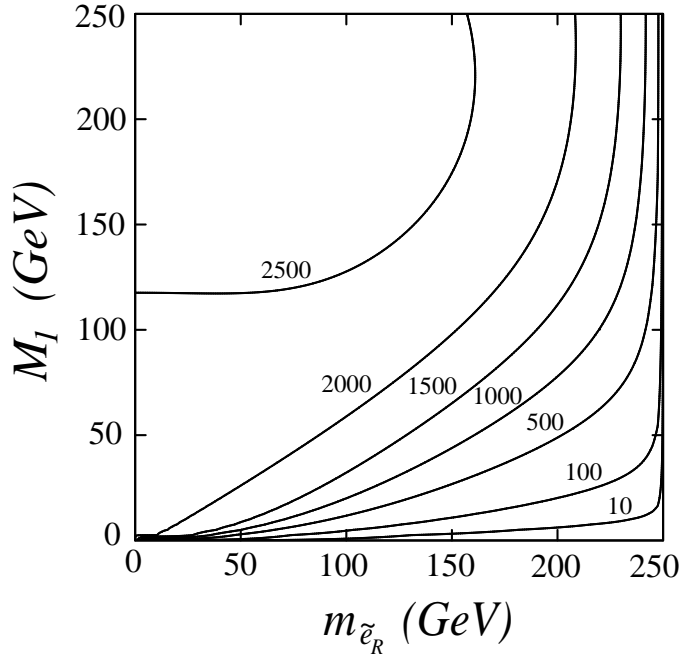,width=0.69\textwidth}}
\caption{Contours of constant $\sigma_R = \sigma (e^-_R e^-_R \to 
\ser^- \ser^-)$ in fb in the ($m_{\ser}$, 
$M_1$) plane for $\protect\sqrt{s}=500$ GeV.
\label{xsection}}
\end{figure}

The total cross section $\sigma_R = \sigma(e_R^- e_R^- \to
\ser^- \ser^-)$ at a 500 GeV $\emem$ collider is shown in
Fig.~\ref{xsection}. 
For a wide range of the selectron
mass and Bino mass\footnote{We assume that we are in the
gaugino region, $\LSP \approx \tilde{B}$, $\mLSP \approx M_1$.
Neutralino mixings will be discussed later.}, ($\mser$ not
too close to the threshold and $M_1$ not too small,) the
cross section is quite large and is on the order of $\sim 2000$ fb. 
Assuming one year running at luminosity ${\cal L} \sim 20
\mbox{ fb}^{-1}/\mbox{yr}$, we expect $\sim 40000$ events, yielding a
statistical uncertainty of $\sim 0.5\%$ for the cross section.
This can be further improved by longer runs or a larger
luminosity. The major
background is $e^- \nu W^-$ when followed by $W^- \to e^-\nu_e$,
which results from $e_L^-$ contamination in the $e^-_R$ polarized
beams. If both beams are 90\%
right-polarized, {\em i.e.}, if only 10\% of the electrons in each
beam are left-handed, the background is reduced to 12 fb~\cite{Cuypers}.
In principle
these backgrounds are calculable and can be subtracted, so the induced
uncertainty in $\sigma_R$ should be negligible.

\begin{figure}
\centerline{\psfig{file=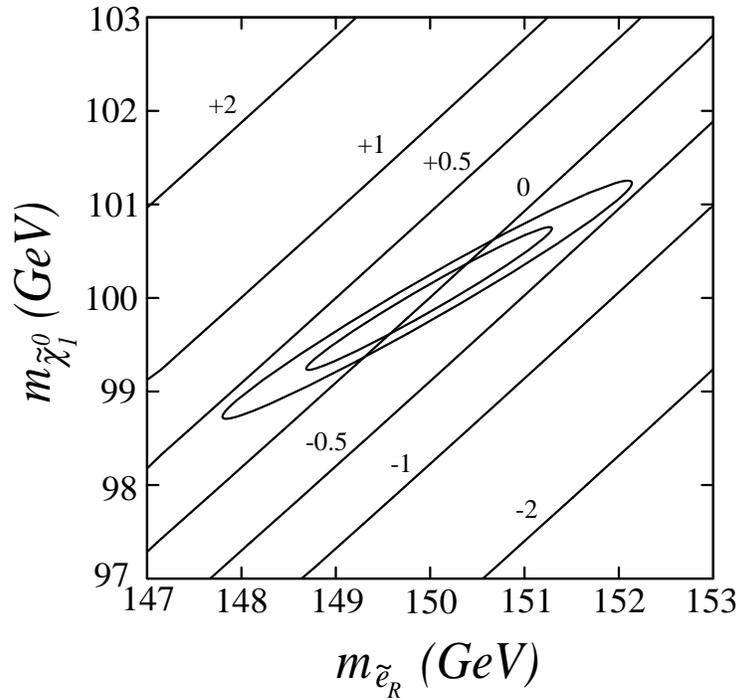,width=0.75\textwidth}}
\caption{The allowed regions, ``uncertainty ellipses,'' of the 
($\mser$, $m_{\LSP}$) plane, determined by measurements of
the end points of final state electron energy distributions with
uncertainties $\Delta E= 0.3$ GeV and 0.5 GeV.  The underlying central
values are $(m_{\ser}, m_{\LSP}) = (150 \mbox{ GeV}, 100
\mbox{ GeV})$, and $\protect\sqrt{s}=500$ GeV.  We also superimpose 
contours (in percent) of the fractional variation of $\sigma_R$ with
respect to its value at the underlying parameters.
\label{ellipse}}
\end{figure}

There are also theoretical systematic errors coming from uncertainties
in the $\ser$ and $\LSP$ masses. the $\ser$ and $\LSP$ masses are
constrained by electron energy distribution endpoints of the electrons
from $\ser$ decays. The resulting allowed masses are positively 
correlated and lie in an ellipse-like region in the $\mser-\mLSP$
plane~\cite{JLCI}. For example, for $\mser=150$ GeV, $\mLSP=100$ GeV,
the typical allowed regions from a year's worth of data are given by
the ellipses in Fig.~\ref{ellipse}. 
On the other hand, the cross
section increases as $M_1$ increases (in the regions in which we are 
interested) because the chirality-flipping mass insertion is needed,
but decreases as $\mser$ increases. As a result, the contours of constant 
cross section run nearly parallel to the major axes of the ellipses,
resulting in a very small error in the total cross section from
the uncertainties of $\mser$ and $\mLSP$, about only 0.3\% in 
this example. In constrast, at the $\epem$ collider, the cross
section decreases as either $\mser$ or $M_1$ increases. The contours
of constant cross section will then run roughly perpendicular to the major
axes of the ellipses, resulting in larger theoretical systematic
errors.

\begin{figure}
\centerline{\psfig{file=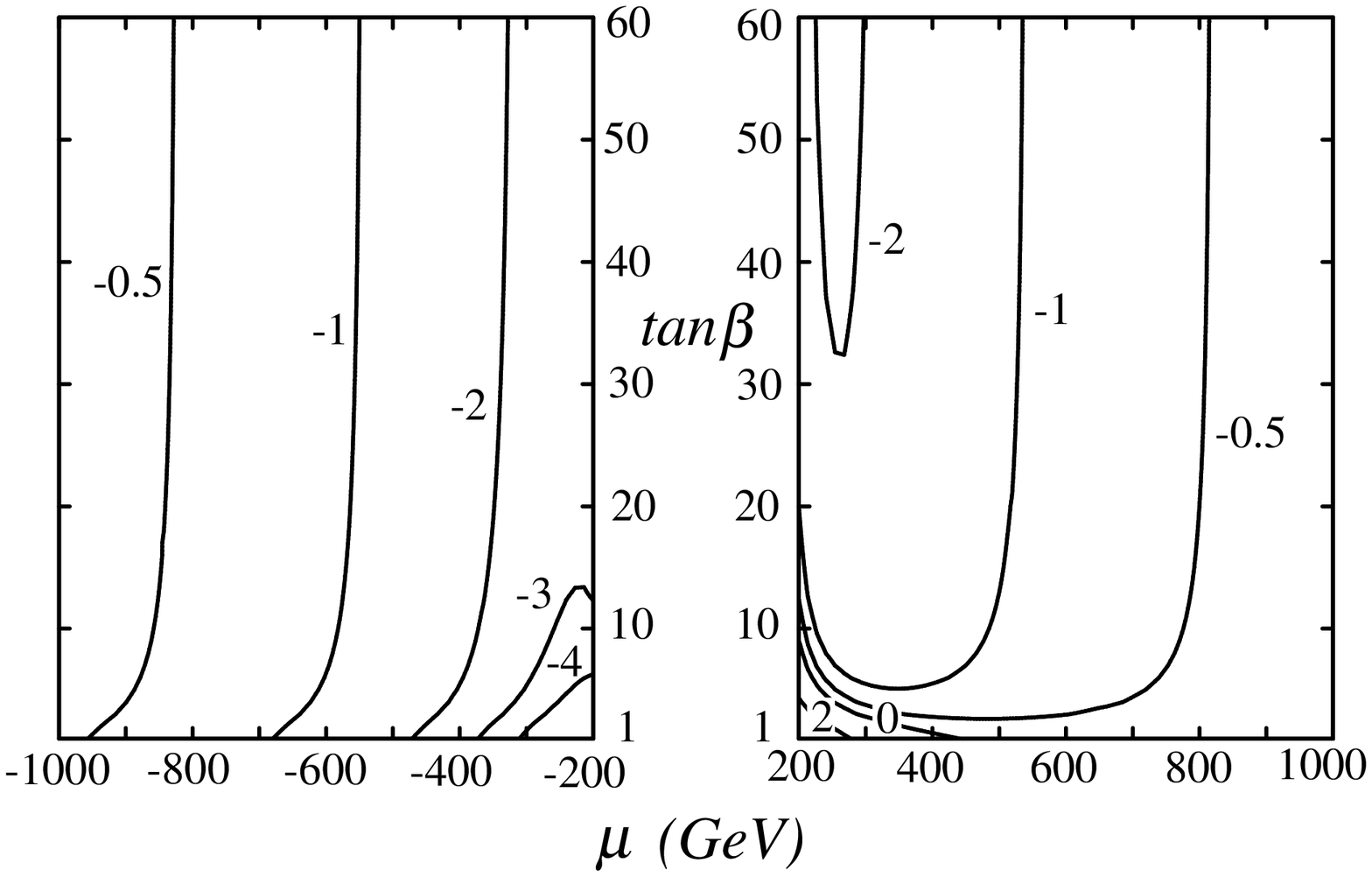,width=0.9\textwidth}}
\caption{The fractional variation in $\sigma_R$ (in percent) in the 
($\mu$, $\tan\beta$) plane, with respect to the $\mu\to \infty$ limit,
for $(m_{\ser}, m_{\LSP})= (150 \mbox{ GeV}, 100\mbox{ GeV})$,
with $\protect\sqrt{s}=500$ GeV.  $M_2$ is assumed to be $2M_1$, and
for each point in the plane, their values are fixed by $m_{\LSP}$.
\label{mutanbeta}}
\end{figure}

Up to now we have assumed that
the lightest neutralino is pure Bino. This is only true in the limit
of $|\mu| \to \infty$. For finite $\mu$, neutralino mixings will 
appear at the $e_R-\ser-\neutralino_i$ vertices, so the cross section will
also depend on other SUSY parameters in the neutralino mass matrix,
$M_2$, $\mu$, and $\tan \beta$, in addition to $M_1$. The $M_2$
dependence is very small because $\tilde{B}$ and $\tilde{W}^3$
only mix indirectly, and hence can be neglected. The $\mu$ and 
$\tan\beta$ dependences are shown in Fig.~\ref{mutanbeta}.
The variation in the cross section is less than 1\% for $|\mu|
\gtrsim 500\sim 600$ GeV, but can be up to 2--4\% for smaller
$|\mu|$. Therefore, some information about $\mu$ and $\tan\beta$
is required to calculate $\sigma_R$ at high precision.
Such information may be obtained from other processes in 
different colliders, for example, $\neutralino_1 \neutralino_3$
production or chargino production in $\epem$ collisions.
Energies of $\sqrt{s}\sim 1$ TeV,
if available, will therefore allow either a
determination of $\mu$ or a sufficiently high lower bound on $\mu$ for
us to obtain a precise prediction of $\sigma_R$ so that $h_1$ can be
extracted with small uncertainties.
Extra mixings at the $e_R-\ser-\neutralino_i$ vertices may exist if
lepton flavor is not conserved, {\it i.e.}, the lepton and slepton
mass matrices are not diagonalized in the same basis. This may 
also cause some uncertainties. However, lepton flavor violation
also can be studied at the same time. For instance, Ref.~11 shows
that a mixing angle between the first and second generations of
order $\sin \theta_{12}\sim 0.02$ will be probed at the 5$\sigma$
level in $\emem$ collisions. With such a precision, the induced
uncertainty in $h_1$ is negligible.

Finally, many of the above considerations apply also to left-handed
selectrons.  If kinematically accessible, their production cross
section $\sigma_L$ at $\emem$ colliders may also be used to precisely
measure gaugino couplings, since the $\sel^-
\sel^-$ pair production cross section receives contributions
from both $t$-channel $\tilde{B}$ and $\tilde{W}^3$ exchange, and
hence depends on both $h_1$ and $h_2$. For equivalent mass selectrons,
$\sigma_L$ is generally even larger than $\sigma_R$.  Note also that
$\sel$ and $\ser$ production may be separated either
by beam polarization, or, if the selectrons are sufficiently
nondegenerate, by kinematics~\cite{JLCI} or by running below the
higher production thresholds.  If the $\tilde{\chi}^{\pm}_1$ and
$\neutralino_2$ decay channels are not open, the only decay is
$\sel^-\to e^-\LSP$ and we will have a large clean sample of
events for precision studies.  However, in general, the decay patterns
may complicate the analysis.  The cross section also depends strongly
on $m_{\neutralino_2}$ (in the gaugino region), which could be
measured either directly from $\neutralino_2 \neutralino_2$
production in $\epem$ collisions, or indirectly by measuring $M_1$,
$M_2$, $\mu$ and $\tan\beta$ from chargino and $\LSP$ properties. In
the end, a measurement of $\sigma_L$ bounds a certain combination of
$h_1$ and $h_2$. Under the assumption that the heavy sparticles are
fairly degenerate, the deviations $\widetilde{U}_1$ and $\widetilde{U}_2$ are
related and determined by the same heavy scale $M$, and so $\sigma_L$
also provides a probe of the heavy scale $M$, which, in fact, is
generically more sensitive, since $\widetilde{U}_2 > \widetilde{U}_1$ 
in most models.  Of course, in the event that both $\ser$ and
$\sel$ are studied, both $\widetilde{U}_1$ and $\widetilde{U}_2$ may be
determined, and we may check that their implications for the heavy
scale $M$ are consistent or find evidence for nondegeneracies in the
heavy sector.

\section{Conclusions}

In summary, we have seen that the $\emem$ collider may provide
the most precise determination of the superparticle couplings.
Of course, it is important that the experimental systematic errors from 
uncertainties in various collider parameters, including the beam 
energy, luminosity, and so on, have to be controlled in order
to obtain the high precision measurements. From this workshop,
it seems that required precision of these collider parameters should 
be able to be achieved. The implications of such precision measurements
of super-oblique parameters depend on the scenario realized in nature.
At the first step, it provides a stringent test of supersymmetry.
If some number of superpartners are not yet discovered, bounds on the
super-oblique parameters may lead to bounds on the mass scale of the
heavy particles. If, on the other hand, all superpartners of the 
standard model
particles are found, the consistency of all super-oblique parameters
with the predicted values will be an important check of the 
supersymmetric model with minimal field content.  
If instead deviations are found, such measurements will provide
exciting evidence for new exotic sectors with highly split multiplets
not far from the weak scale~\cite{CFP1}. These insights
could also provide a target for future superparticle searches, and could
play an important role in evaluating future proposals for colliders
with even higher energies.


\end{document}